\documentstyle[aps,12pt]{revtex}

\begin{document}
\author{E.N. Glass\thanks{%
Permanent address: Physics Department, University of Windsor, Ontario N9B
3P4, Canada} and J.P. Krisch}
\address{Department of Physics, University of Michigan, Ann Arbor, Michgan 48109}
\date{December 6, 1997}
\title{Radiation and String Atmosphere for Relativistic Stars}
\maketitle

\begin{abstract}
\\ \\ We extend the Vaidya radiating metric to include both a radiation
field and a string fluid. Assuming diffusive transport for the string fluid,
we find new analytic solutions of Einstein's field equations. Our new
solutions represent an extention of Xanthopoulos superposition.\newpage\ 
\end{abstract}

Vacuum solutions of Einstein's field equations have played an important role
in our understanding of curvature effects and relativistic behavior even
though it is obvious that real stars do not sit in vacuum but have particle
and radiation atmospheres. Quantum effects allow atmospheres to be added to
classical vacuums; for example black holes are associated with atmospheres
of Hawking radiation \cite{thooft},\cite{hiscock}. In addition to their
intrinsic value as exact solutions, vacuum solutions in general relativity
are approximate string theory solutions for curvature small compared to the
Planck scale \cite{garetal}. The intense level of activity in string theory
has lead to the idea that many of the classic vacuum scenarios, such as the
static Schwarzschild point mass/black hole, may have atmospheres composed of
a fluid or field of strings \cite{par&vis}. One of the well known examples
of radiation atmospheres is the Vaidya metric \cite{vaidya}, generated from
the vacuum Schwarzschild solution by allowing the Schwarzschild mass to be a
function of retarded time. The resulting stress-energy content describes
outgoing short-wavelength photons.

In this letter we extend the Vaidya metric by allowing the mass to be a
function of both retarded time and distance along the outgoing null
geodesics. The effect of this extension is to create two fluids outside the
star, the original null fluid and a new fluid composed of strings. Given the
recent links \cite{kar},\cite{larsen} between black holes and string
theories, this result is of interest by itself. It is additionally
interesting since our new analytic solutions for the mass function allow the
metric to be written as a superposition of a string fluid and vacuum
Schwarzschild. We have thereby extended Xanthopolous superposition \cite{xan}%
.

The string fluid tension and density depend on spatial derivatives of the
mass function. Assuming a specific model for propagation of the density
allows the generation of new densities and hence new mass functions. We
choose to propagate the density diffusively as a particular example of mass
transport.

Our sign conventions are $2A_{\nu ;[\alpha \beta ]}=A_\mu R_{\ \ \nu \alpha
\beta }^\mu ,$ $R_{\mu \nu }=R_{\ \mu \nu \alpha }^\alpha $, and metric
signature (+,-,-,-). Greek indices range over (0,1,2,3) = ($u,r,\vartheta
,\varphi $). $\dot{m}$ abbreviates $\partial m/\partial u$, $m^{\prime }$
abbreviates $\partial m/\partial r$, and $m^{\prime \prime }$ represents $%
\partial ^2m/\partial r^2$. Overhead carets denote unit vectors.

The spacetime metric covering the region exterior to a spherical star is
given by 
\begin{equation}
ds^2=Adu^2+2dudr-r^2(d\vartheta ^2+\text{sin}^2\vartheta d\varphi ^2)
\label{met1}
\end{equation}
where $A=1-2m(u,r)/r$. Initially $m(u,r)=m_0$ provides the vacuum
Schwarzschild solution in the region $r>2m_0$. At later times $m(u,r)$
admits a two-fluid description of diffusing matter and outward flowing
short-wavelength photons (sometimes called a ''null fluid''). Metric (\ref
{met1}) is spherically symmetric and given in retarded time coordinate $u$.
With the use of a Newman-Penrose null tetrad the Einstein tensor is computed
from (\ref{met1}) and given by 
\begin{eqnarray}
G_{\mu \nu } &=&-2\Phi _{11}(l_\mu n_\nu +n_\mu l_\nu +m_\mu \bar{m}_\nu +%
\bar{m}_\mu m_\nu )  \label{ein1} \\
&&-2\Phi _{22}l_\mu l_\nu -6\Lambda g_{\mu \nu }.  \nonumber
\end{eqnarray}
Here the null tetrad components of the Ricci tensor are 
\begin{mathletters}
\label{ein2}
\begin{eqnarray}
\Phi _{11} &=&-(rm^{\prime \prime }-2m^{\prime })/(4r^2),  \label{ein1a} \\
\Phi _{22} &=&-\dot{m}/r^2,  \label{ein1b} \\
\Lambda &=&R/24=(rm^{\prime \prime }+2m^{\prime })/(12r^2).  \label{ein1c}
\end{eqnarray}
The metric is Petrov type D with $l_\mu $ and $n_\mu $ principal null
vectors, $l_\mu $ geodesic, and 
\end{mathletters}
\begin{mathletters}
\label{ntet}
\begin{eqnarray}
l_\mu dx^\mu &=&du,  \label{nteta} \\
n_\mu dx^\mu &=&(A/2)du+dr,  \label{ntetb} \\
m_\mu dx^\mu &=&-(r/\surd 2)(d\vartheta +i\ \text{sin}\vartheta d\varphi ).
\label{ntetc}
\end{eqnarray}
In order to clearly see the two-fluid description we introduce a timelike
unit velocity vector $\hat{v}^\mu $ and three unit spacelike vectors $\hat{r}%
^\mu $, $\hat{\vartheta}^\mu $, $\hat{\varphi}^\mu $ such that 
\end{mathletters}
\[
g_{\mu \nu }=\hat{v}_\mu \hat{v}_\nu -\hat{r}_\mu \hat{r}_\nu -\hat{\vartheta%
}_\mu \hat{\vartheta}_\nu -\hat{\varphi}_\mu \hat{\varphi}_\nu . 
\]
The unit vectors are defined by 
\begin{mathletters}
\label{vtet}
\begin{eqnarray}
\hat{v}_\mu dx^\mu &=&A^{1/2}du+A^{-1/2}dr,  \label{vteta} \\
\hat{r}_\mu dx^\mu &=&A^{-1/2}dr,  \label{vtetb} \\
\hat{\vartheta}_\mu dx^\mu &=&rd\vartheta ,  \label{vtetc} \\
\hat{\varphi}_\mu dx^\mu &=&r\text{sin}\vartheta d\varphi .  \label{vtetd}
\end{eqnarray}
The Einstein tensor (\ref{ein1}) can be written as a two-fluid system: 
\end{mathletters}
\begin{eqnarray}
G_{\mu \nu } &=&(2\dot{m}/r^2)l_\mu l_\nu -(2m^{\prime }/r^2)(\hat{v}_\mu 
\hat{v}_\nu -\hat{r}_\mu \hat{r}_\nu )  \label{ein3} \\
&&+(m^{\prime \prime }/r)(\hat{\vartheta}_\mu \hat{\vartheta}_\nu +\hat{%
\varphi}_\mu \hat{\varphi}_\nu ).  \nonumber
\end{eqnarray}
Spherical symmetry allows the function $m(u,r)$ to be identified as the mass
within two-surfaces of constant $u$ and $r$, and invariantly defined from
the sectional curvature of those surfaces: 
\begin{equation}
-2m/r^3=R_{\alpha \beta \mu \nu }\hat{\vartheta}^\alpha \hat{\varphi}^\beta 
\hat{\vartheta}^\mu \hat{\varphi}^\nu .  \label{mdef}
\end{equation}

The string bivector is defined by 
\[
\Sigma ^{\mu \nu }=\epsilon ^{BC}\frac{\partial x^\mu }{\partial x^B}\frac{%
\partial x^\nu }{\partial x^C},\ \ \ (B,C)=(0,1)\ or\ (2,3). 
\]
Spherical symmetry requires that the string bivector have a world-sheet in
either the ($u,r$) or ($\vartheta ,\varphi $) plane. We require that the
world-sheets be timelike, i.e. $\gamma :=\frac 12\Sigma ^{\mu \nu }\Sigma
_{\mu \nu }<0$, and so only the $\Sigma _{ur}$ component is non-zero. Here $%
\gamma =-1$. It is useful to write $\Sigma ^{\mu \nu }$ in terms of unit
vectors 
\begin{equation}
\Sigma ^{\mu \nu }=\hat{r}^\mu \hat{v}^\nu -\hat{v}^\mu \hat{r}^\nu ,
\label{stringvec}
\end{equation}
and so $\Sigma ^{\mu \alpha }\Sigma _\alpha ^{\ \nu }=\hat{v}^\mu \hat{v}%
^\nu -\hat{r}^\mu \hat{r}^\nu $. We follow Letelier \cite{let1},\cite{let2}
and write a string energy-momentum tensor by analogy with one for a perfect
fluid. The string energy-momentum is given by 
\[
T_{\mu \nu }^{string}=\rho (-\gamma )^{1/2}\hat{\Sigma}_\mu ^{\ \alpha }\hat{%
\Sigma}_{\alpha \nu }-p_{\perp }H_{\mu \nu }, 
\]
where $H_{\ \nu }^\mu =\delta _{\ \nu }^\mu -\hat{\Sigma}^{\mu \alpha }\hat{%
\Sigma}_{\alpha \nu }$, $\ H_{\ \nu }^\mu \hat{\Sigma}^{\nu \beta }=0$. We
have written $\hat{\Sigma}^{\mu \nu }:=(-\gamma )^{-1/2}\Sigma ^{\mu \nu }$,
so that $\hat{\Sigma}^{\mu \nu }$ is invariant under reparameterizations of
the world-sheets \cite{let1}. Einstein's field equations, $G_{\mu \nu
}=-8\pi T_{\mu \nu }$, allow the matter portion of (\ref{ein3}) to be
identified as a string fluid: 
\begin{equation}
T_{\mu \nu }=\psi l_\mu l_\nu +\rho \hat{v}_\mu \hat{v}_\nu +p_r\hat{r}_\mu 
\hat{r}_\nu +p_{\perp }(\hat{\vartheta}_\mu \hat{\vartheta}_\nu +\hat{\varphi%
}_\mu \hat{\varphi}_\nu ).  \label{energymom}
\end{equation}
Thus 
\begin{mathletters}
\label{emom}
\begin{eqnarray}
4\pi \psi &=&-\dot{m}/r^2,  \label{psi} \\
4\pi \rho &=&-4\pi p_r=m^{\prime }/r^2,  \label{rho} \\
8\pi p_{\perp } &=&-m^{\prime \prime }/r.  \label{pperp}
\end{eqnarray}
The equation of motion $T_{\ ;\nu }^{\mu \nu }=0$ is identically satisfied
for the components of $T_{\mu \nu }$ given in Eqs.(\ref{emom}).

As an example of mass transport we assume the strings diffuse and that
string diffusion is like point particle diffusion where the number density
diffuses from higher numbers to lower according to 
\end{mathletters}
\begin{equation}
\partial _un={\cal D}\ \nabla ^2n.  \label{ndifu}
\end{equation}
$\nabla ^2=r^{-2}(\partial /\partial r)r^2(\partial /\partial r)$, and $%
{\cal D}$ is the positive coefficient of self-diffusion. Classical transport
theory derives the diffusion equation by starting with Fick's law 
\begin{equation}
\vec{J}_{(n)}=-{\cal D}\vec{\nabla}n  \label{fick}
\end{equation}
where $\vec{\nabla}$ is a purely spatial gradient. Then 4-current
conservation $J_{(n);\mu }^\mu =0,$ where 
\begin{eqnarray}
J_{(n)}^\mu \partial _\mu &=&(n,\vec{J}_{(n)})  \label{jvec} \\
&=&n\partial _u-{\cal D}(\partial n/\partial r)\partial _r,  \nonumber
\end{eqnarray}
yields the diffusion equation (\ref{ndifu}). We label the 4-current $J_{(n)}$
to indicate $n$ diffusion but we could have also written $J_{(\rho )}$ since
the string number density and string fluid density must be related by $\rho
=M_sn$ where $M_s$ is the constant mass of the string species. $M_s$ must be
a multiple of the Planck mass since it is only over Planck length scales
that point particles resolve into strings.

By rewriting the $T_{\mu \nu }$ components (\ref{psi}) and (\ref{rho}) as $%
\dot{m}=-4\pi r^2\psi $ and $m^{\prime }=4\pi r^2\rho ,$ we can write the
integrability condition for $m$ as 
\begin{equation}
\dot{\rho}+r^{-2}\partial _r(r^2\psi )=0.  \label{rhocontinuity}
\end{equation}
If we compare the diffusion equation (\ref{ndifu}) ($n$ replaced by $\rho $) 
\begin{equation}
\dot{\rho}={\cal D}\ r^{-2}\partial /\partial r(r^2\partial \rho /\partial r)
\label{rhodifu}
\end{equation}
with $\dot{\rho}$ in Eq.(\ref{rhocontinuity}) we obtain 
\begin{equation}
\dot{m}=4\pi {\cal D}\ r^2\partial \rho /\partial r.  \label{mdot2}
\end{equation}
Thus solving the diffusion equation for $\rho $ and then integrating those
solutions to obtain $m$ provides exact Einstein solutions for diffusing
string fluids.

There are many analytic solutions of (\ref{rhodifu}) and three of them are 
\begin{mathletters}
\label{solns}
\begin{eqnarray}
\rho &=&\rho _0+k_1/r,  \label{solna} \\
\rho &=&\rho _0+k_3u^{-3/2}\text{exp}[-r^2/(4{\cal D}u)],  \label{solnb} \\
\rho &=&\rho _0+(k_4/r)\{1+(\sqrt{\pi }/2)\text{erf}[r(4{\cal D}u)^{-1/2}]\}.
\label{solnc}
\end{eqnarray}

Upon integrating $m^{\prime }=4\pi r^2\rho $ and $\dot{m}=4\pi {\cal D}\
r^2\partial \rho /\partial r$ we obtain the following masses, listed
consecutively, for the densities above: 
\end{mathletters}
\begin{mathletters}
\label{masses}
\begin{eqnarray}
m(u,r) &=&m_0+(4\pi /3)r^3\rho _0+2\pi k_1(r^2-2{\cal D}u).  \label{mone} \\
m(u,r) &=&m_0+(4\pi /3)r^3\rho _0+16\pi k_3{\cal D}^{3/2}[-\eta \text{exp}%
(-\eta ^2)+(\sqrt{\pi }/2)\text{erf}(\eta )].  \label{mtwo} \\
m(u,r) &=&m_0+(4\pi /3)r^3\rho _0  \label{mthree} \\
&&+2\pi r^2k_4\{(1-\frac 1{2\eta ^2})+[(\sqrt{\pi }/2)(1-\frac 1{2\eta ^2})%
\text{erf}(\eta )+(2\eta )^{-1}\text{exp}(-\eta ^2)]\}.  \nonumber
\end{eqnarray}
where $\eta :=r\,(4{\cal D}u)^{-1/2}$.

It is clear that metric (\ref{met1}) can be written in Kerr-Schild form as $%
\eta _{\mu \nu }-[2m(u,r)/r]l_\mu l_\nu $. When $m(u,r)=m_0$ for the
Schwarzschild solution, $(2m_0/r)l_\mu l_\nu $ solves the vacuum field
equations linearized about flat space. This was Xanthopoulos' original
superposition. His generalization \cite{xan} has $\hat{g}_{\mu \nu }=g_{\mu
\nu }+H(x^\mu )l_\mu l_\nu $ with $H(x^\mu )l_\mu l_\nu $ required to solve
the vacuum field equations linearized about $g_{\mu \nu }$. Referring to the
mass solutions above, we can write $1-2m(u,r)/r=1-2m_0/r-2\tilde{m}/r$.
Metric (\ref{met1}) then has the form $\eta _{\mu \nu }-(2m_0/r)l_\mu l_\nu
-(2\tilde{m}/r)l_\mu l_\nu $ which is clearly 
\end{mathletters}
\begin{equation}
\hat{g}_{\mu \nu }=g_{\mu \nu }^{Sch}-(2\tilde{m}/r)l_\mu l_\nu .
\label{supermet}
\end{equation}
For the Vaidya metric, with $\tilde{m}(u)$, the field equations linearized
about $g_{\mu \nu }^{Sch}$ yield $G_{\mu \nu }^{(1)}=(d\tilde{m}%
/du)r^{-2}l_\mu l_\nu $ \cite{glass} which is not a vacuum solution and so $%
G_{\mu \nu }^{(1)}$ computed about Schwarzschild with $\tilde{m}(u,r)$ is 
{\it a fortiori} not zero. Since $\hat{g}_{\mu \nu }$ is an exact solution
for a string fluid and $g_{\mu \nu }^{Sch}$ is an exact vacuum solution, we
have extended Xanthopolous' generalization.\\\\E.N. Glass was partially
supported by an NSERC of Canada grant. Computations were verified using
MapleV.4 (Waterloo Maple Software, Waterloo, Ontario) and GRTensorII rel
1.59 (P. Musgrave, D.Pollney, and K. Lake, Queens University, Kingston,
Ontario).


\begin{references}
\bibitem{thooft}  G. 't Hooft, ''The Self-Screening Hawking Atmosphere'',
presented at Strings97, Amsterdam, 1997\ [gr-qc/9706058].

\bibitem{hiscock}  W.A. Hiscock, Phys. Rev. {\bf D15}, 3054 (1977).

\bibitem{garetal}  D. Garfinkle, G. Horowitz, and A. Strominger, Phys. Rev. 
{\bf D43}, 3140 (1991).

\bibitem{par&vis}  R. Parthasarathy and K.S. Viswananthan, Phys. Lett. {\bf %
B400,} 27 (1992).

\bibitem{vaidya}  P.C. Vaidya, Nature {\bf 171,} 260 (1953).

\bibitem{kar}  S. Kar, Phys. Rev. {\bf D55}, 4872 (1997).

\bibitem{larsen}  F. Larsen, Phys. Rev. {\bf D56}, 1005 (1997).

\bibitem{xan}  K.E. Mastronikola and B.C. Xanthopoulos, Class. Quantum Grav. 
{\bf 6}, 1613 (1989).

\bibitem{let1}  P.S. Letelier, Phys. Rev. {\bf D20}, 1294 (1979).

\bibitem{let2}  P.S. Letelier, Nuov. Cim. {\bf 63B}, 519 (1981).

\bibitem{glass}  E.N. Glass, Phys. Rev. {\bf D47}, 474 (1993).
\end{references}
\end{document}